\documentclass[pra, 12 pt]{revtex4}
\usepackage[english]{babel}
\usepackage{amsmath}
\usepackage{epsfig}

\begin{document}
\title{Anomalous Kinetics in Velocity Space: equations and models}
\author{S.A. Trigger}
\address{Joint\, Institute\, for\, High\, Temperatures, Russian\, Academy\,
of\, Sciences, 13/19, Izhorskaia Str., Moscow\, 125412, Russia;\\
email:\,satron@mail.ru}

\begin{abstract}

Equation for anomalous diffusion in momentum space, recently
obtained in [1], is solved for the stationary and non-stationary
cases on basis of the appropriate probability transition function
(PTF). Consideration of diffusion for heavy particles in a gas of
the light particles can be essentially simplified due to small
ratio of the masses of the particles. General equation for the
distribution of the light particles, shifted in velocity space, is
also derived. For the case of anomalous diffusion in momentum
space the closed equation is formulated for the Fourier-component
of the momentum distribution function. The effective friction and
diffusion coefficients are found also for the shifted
distribution. If the appropriate integrals are finite the
equations derived in the paper are applicable for both cases: the
PT-function with the long tails and the short range PT-functions
in momentum space. In the last case the results are equivalent to
the Fokker-Planck equation. Practically the new results of this
paper are applicable
to strongly non-equilibrium physical systems.\\

PACS number(s): 52.27.Lw, 52.20.Hv, 05.40.-a, 05.40.Fb

\end{abstract}

\maketitle

\section{Introduction}
Interest in anomalous diffusion is conditioned by a large variety
of applications: semiconductors, polymers, some granular systems,
plasmas in specific conditions, various objects in biological
systems, physical-chemical systems, et cetera.

Many concrete problems of anomalous diffusion in coordinate space
have been solved on the basis of equations with the fractional
derivatives. Recently in [2] the new approach to anomalous
diffusion in coordinate space has been formulated. This approach
simplifies the problem and at the same time permits to consider
the more complicated kernels (PTF in coordinate space). In the
present paper, as well as in [1], we apply the similar approach to
solve the problem in momentum space.

The deviation from the linear in time $<r^2(t)>\sim t$ dependence
of the mean square displacement have been experimentally observed,
in particular, under essentially non-equilibrium conditions or for
some disordered systems. The average square separation of a pair
of particles passively moving in a turbulent flow grows, according
to Richardson's law, with the third power of time [3]. For
diffusion typical for glasses and related complex systems [4] the
observed time dependence is slower than linear. These two types of
anomalous diffusion obviously are characterized as superdiffusion
$<r^2(t)>\sim t^\alpha$ $(\alpha>1)$ and subdiffusion $(\alpha<1)$
[5]. For a description of these two diffusion regimes a number of
effective models and methods have been suggested. The continuous
time random walk (CTRW) model of Scher and Montroll [6], leading
to strongly subdiffusion behavior, provides a basis for
understanding photoconductivity in strongly disordered and glassy
semiconductors. The Levy-flight model [7], leading to
superdiffusion, describes various phenomena as self-diffusion in
micelle systems [8], reaction and transport in polymer systems [9]
and is applicable even to the stochastic description of financial
market indices [10]. For both cases the so-called fractional
differential equations in coordinate and time spaces are applied
as an effective approach [11].

However, recently a more general approach has been suggested in
[2], [12], which avoid the fractional differentiation, reproduce
the results of the standard fractional differentiation method,
when the last one is applicable, and permit to describe the more
complicated cases of anomalous diffusion processes. In [13] these
approach has been applied also to the diffusion in the
time-dependent external field.

In this paper the problem of anomalous diffusion in the momentum
(velocity) space will be considered. In spite of formal
similarity, diffusion in the momentum space is very different
physically from the coordinate space diffusion. It is clear
already because the momentum conservation, which take place in the
momentum space has no analogy in the coordinate space.

Some aspects of the anomalous diffusion in the velocity space have
been investigated for the last decade in a few papers [14-18]. On
the whole, comparing with the anomalous diffusion in coordinate
space, the anomalous diffusion in velocity space is weakly
studied. The consequent way to describe the anomalous diffusion in
the velocity space is, according to our knowledge, still absent.

In this paper the new kinetic equation for anomalous diffusion in
velocity space is derived (see also [1]) on the basis of the
appropriate expansion of PTF (in the spirit of the approach
suggested in [2] for the diffusion in coordinate space) and some
particular problems are investigated on this basis.

The diffusion in velocity space for the cases of normal and
anomalous behavior of the PT function is presented in the Section
II. Starting from the argumentation based on the Boltzmann type of
the PTF, we derive the new kinetic equation, which in fact can be
applied to the wide class of the PTF functions. The particular
cases of anomalous diffusion for hard spheres collisions with the
specific power-type prescribed distribution function of the light
particles is analyzed in the Section III. The universal character
of anomalous diffusion in velocity space is absent for this case.
But for the general case of the power-type PTF with the different
powers (which are not connected in advance) the universality takes
a place. For this case the universal limitations for the existence
of anomalous diffusion are found. In the Section IV the Boltzmann
type equation is used to consider influence of the drift of the
light particles on the PTF function and on the opportunity for
anomalous transport of the heavy component.

\section{Diffusion in the velocity space on the basis of
a master-type equation}

Let us consider now the main problem formulated in the
introduction, namely, diffusion in velocity space ($V$-space) on
the basis of the respective master equation, which describes the
balance of grains coming in and out the point $p$ at the moment
$t$. The structure of this equation is formally similar to the
master equation Eq.~(\ref{DC2}) in the coordinate space
\begin{equation}
\frac{df_g({\bf p},t)}{dt} = \int d{\bf q} \left\{W ({\bf q, p+q})
f_g({\bf p+q}, t) - W ({\bf q, p}) f_g({\bf p},t) \right\}.
\label{DC2b}
\end{equation}
Of course, for coordinate space there is no conservation law,
similar to that in the momentum space. The probability transition
$W({\bf p, p'})$ describes the probability for a grain with
momentum ${\bf p'}$ (point ${\bf p'}$) to transfer from this point
${\bf p'}$ to the point ${\bf p}$ per unit time. The momentum
transferring is equal ${\bf q= p'- p}$. Assuming in the beginning
that the characteristic displacements are small one may expand
Eq.~(\ref{DC2})  and arrive at the Fokker-Planck form of the
equation for the density distribution $f_g({\bf p},t)$
\begin{equation}
\frac{df_g({\bf p},t)}{dt} = \frac {\partial}{\partial p_\alpha}
\left[ A_\alpha ({\bf p}) f_g({\bf p},t) + \frac{\partial}
{\partial p_\beta} \left(B_{\alpha\beta}({\bf p}) f_g({\bf p},t)
\right)\right]. \label{DC3b}
\end{equation}
\begin{equation}
A_\alpha({\bf p}) = \int d^s q q_\alpha W({\bf q, p});\;\;\;\
B_{\alpha\beta}({\bf p})= \frac{1}{2}\int d^s q q_\alpha q_\beta
W({\bf q, p}). \label{DC4b}
\end{equation}
The coefficients $A_\alpha$ and $B_{\alpha \beta}$ describing the
friction force and diffusion, respectively.

Because the velocity of heavy particles is small, the $\bf
p$-dependence of the PTF can be neglected for calculation of the
diffusion, which in this case is constant
$B_{\alpha\beta}=\delta_{\alpha\beta}B$, where B is the integral
\begin{equation}
B = \frac{1}{2s}\int d^s q q^2 W(q). \label{DC6b}
\end{equation}

If to neglect the $\bf p$-dependence of the PTF at all we arrive
to the coefficient $A_\alpha=0$ (while the diffusion coefficient
is constant). This neglecting, as well known is wrong, and the
coefficient $A_\alpha$ for the Fokker-Planck equation can be
determined by use the argument that the stationary distribution
function is Maxwellian. On this way we arrive to the standard form
of the coefficient $MT A_\alpha(p)=p_\alpha B$, which is one of
the forms of Einstein relation. For the systems far from
equilibrium this argument is not acceptable.

To find the coefficients in the kinetic equation, which are
applicable also to slowly decreasing PT functions, let us use a
more general way, based on the difference of the velocities of the
light and heavy particles. For calculation of the function
$A_\alpha$ we have take into account that the function $W(\bf
{q,p})$ is scalar and depends on $q, {\bf q \cdot p}, p$.
Expanding $W(\bf {q,p})$ on $\bf {q\cdot p}$ one arrive to the
approximate representation of the functions $W(\bf {q,p})$ and
$W({\bf q, p+q})$:
\begin{eqnarray}
W({\bf q,p)}\simeq W(q)+\tilde W'(q)({\bf q \cdot p})+
 \frac{1}{2}\tilde W''(q) ({\bf q \cdot p})^2 . \label{DC7b}
\end{eqnarray}
\begin{equation}
W ({\bf q, p+q})\simeq W(q)+\tilde W'(q) \,({\bf q \cdot
p})+\frac{1}{2}\tilde W''(q) ({\bf q \cdot p})^2 +q^2\tilde
W'(q),\label{DC9b}
\end{equation}
where $\tilde W'(q)\equiv \partial W (q, {\bf q \cdot p})
/\partial ({\bf q p})\mid_{{\bf q \cdot p}=0}$ and $\tilde
W''(q)\equiv \partial^2 W (q, {\bf q \cdot p}) /\partial ({\bf q
p})^2 \mid_{{\bf q \cdot p}=0}$.

Then, with the necessary accuracy, $A_\alpha$ equals
\begin{equation}
A_\alpha({\bf p}) = \int d^s q q_\alpha q_\beta p_\beta \tilde
W'(q)= p_\alpha \int d^s q q_\alpha q_\alpha  \tilde
W'(q)=\frac{p_\alpha}{s} \int d^s q q^2  \tilde W'(q)\label{DC10b}
\end{equation}
If for the function $W({\bf q,p)}$ the equality $\tilde W'(q)=
W(q)/ 2 MT$ is fulfilled, then we arrive to the usual Einstein
relation
\begin{equation}
M T A_\alpha({\bf p}) =  p_\alpha B \label{DC11b}
\end{equation}

Let us check this relation for the Boltzmann collisions, which are
described by the PT-function $W({\bf q, p)}=w_B({\bf q, p})$ [12]:
\begin{eqnarray}
w_B({\bf q, p})=\frac{2\pi}{\mu^2 q} \int_{q/2\mu}^\infty du\,u\,
\frac{d \sigma}{do} \left[\arccos \, (1-\frac{q^2}{2\mu^2 u^2}), u
\right] f_b (u^2+ v^2-{\bf q \cdot v} /\mu), \label{DC12b}
\end{eqnarray}
where (${\bf p}=M{\bf v}$) and $d \sigma (\chi,u)/ do\, $, $\mu$ and $f_b$ are
respectively the differential cross-section for scattering, the mass and
distribution function for the light particles. In Eq.~(\ref{DC12b}) we took into account
the approximate equalities for the scattering of the light and heavy particles $q^2 \equiv (\triangle {\bf p})^2=p'^2(1-cos\theta)$ and $\theta\simeq \chi$, where $p'=\mu u$ is the momentum of the light particle before
collision.

For the equilibrium Maxvellian distribution $f_b^0$ the equality $\tilde W'(q)= W(q)/
2 MT$ is evident and we arrive to the usual Fokker-Planck equation
in velocity space with the constant diffusion $D \equiv B /M^2$
and friction $\beta \equiv B/MT=DM/T$ coefficients, which satisfy
the Einstein relation.

For some non-equilibrium situations the PTF can possess a long
tail. In this case we have derive a generalization of the
Fokker-Planck equation in spirit of the above consideration for
the coordinate case, because the diffusion and friction
coefficients in the form Eqs.~(\ref{DC6b}),(\ref{DC10b}) diverge
for large $q$ if the functions have the asymptotic behavior
$W(q)\sim 1/q^\alpha$ with $\alpha\leq s+2$ and (or) $\tilde
W'(q)\sim 1/q^\beta$ with $\beta \leq s+2$.

Let us insert in Eq.~(\ref{DC2b}) the expansions for $W$ (as an
example we choose $s=3$, the arbitrary $s$ can be considered by
the similar way). With necessary accuracy we find
\begin{eqnarray}
\frac{df_g({\bf p},t)}{dt}=\int d{\bf q} \{f_g({\bf p+q}, t)[W(q)+\tilde W'(q) \,({\bf q \cdot p})+\nonumber\\
\frac{1}{2}\tilde W''(q) ({\bf q \cdot p})^2 +q^2 \tilde
W'(q)]-f_g({\bf
p},t)[W(q)+\tilde W'(q) \,({\bf q \cdot p})+\frac{1}{2}\tilde
W''(q) ({\bf q \cdot p})^2]\} \label{DC13b}
\end{eqnarray}

After the Fourier-transformation $f_g ({\bf r})=\int \frac{d{\bf
p}}{(2\pi)^3} exp(i{\bf p r})f_g ({\bf p},t)$ Eq.~(\ref{DC13b})
reads:
\begin{eqnarray}
\frac{df_g({\bf r},t)}{dt} = \int d{\bf q} \{ exp(-i{\bf(q
r)}[W(q)- i \tilde W'(q) \,({\bf
q} \cdot \frac{\partial}{\partial {\bf r}})\nonumber\\
-\frac{1}{2}\tilde W''(q) ({\bf q} \cdot \frac{\partial}{\partial
{\bf r}})^2] - [W(q) - i \tilde W'(q) \,({\bf q} \cdot
\frac{\partial}{\partial {\bf r}})-\frac{1}{2}\tilde W''(q) ({\bf
q} \cdot\frac{\partial}{\partial {\bf r}})^2]\}f_g({\bf r},
t)\label{DC15b}
\end{eqnarray}

We can rewrite this equation as [1]:
\begin{eqnarray}
\frac{df_g({\bf r},t)}{dt} = A(r)f_g ({\bf r})+ B_\alpha
(r)\frac{\partial f_g ({\bf
r},t)}{\partial {\bf r}_\alpha} +C_{\alpha\beta}(r)\frac{\partial^2 f_g ({\bf r},t)}{\partial {\bf r}_\alpha
\partial {\bf
r}_\beta}\label{DC16b}
\end{eqnarray}
where
\begin{eqnarray}
A(r)= \int d{\bf q} [exp(-i{\bf(q r)})-1]W(q)
= 4\pi \int_0^\infty dq q^2 \left[\frac{sin\, (q r)}{qr}-1\right]W(q)
\label{DC17b}
\end{eqnarray}
\begin{eqnarray}
B_\alpha\equiv r_\alpha B(r);\;B(r)=-\frac{i}{r^2} \int d{\bf q}
{\bf q r} [exp(-i{\bf(q r})-1]  \tilde W'(q)= \nonumber\\
\frac{4\pi}{r^2} \int_0^\infty dq q^2 \left[cos\, (q r)-\frac{sin
(q r)}{q r}\right]\tilde W'(q) \label{DC19b}
\end{eqnarray}
\begin{eqnarray}
C_{\alpha\beta} (r)\equiv r_\alpha r_\beta C(r)= -\frac{1}{2}\int
d{\bf q} q_\alpha q_\beta [exp(-i{\bf(q r})-1]  \tilde
W''(q)\label{DC20b}
\end{eqnarray}
\begin{eqnarray}
C(r )=-\frac{1}{2 r^4} \int d{\bf q} {\bf (q r)^2} [exp(-i({\bf q
r})-1] \tilde W''(q)=\nonumber\\\frac{2\pi}{r^2} \int_0^\infty dq
q^4 \left[\frac{2 sin (q r)}{q^3 r^3}-\frac{2 cos\, (q r)}{q^2
r^2}-\frac{sin (qr)}{qr}+\frac{1}{3}\right]\tilde W''(q)
\label{DC21b}
\end{eqnarray}

For the isotropic function $f({\bf r})=f(r)$ one can rewrite
Eq.~(\ref{DC16b}) in the form
\begin{eqnarray}
\frac{df_g(r,t)}{dt} = A (r) f_g (r)+ B(r)r \frac{\partial f_g (r,t)}{\partial
r} +C(r) r^2 \frac{\partial^2 f_g(r,t)}{\partial r^2} \label{DC22b}
\end{eqnarray}

For the case of strongly decreasing PDF the exponent under the
integrals for the functions $A(r)$, $B(r)$ and $C(r)$ can be
expanded:
\begin{eqnarray}
A(r)\simeq=-\frac{r^2}{6}\int d{\bf q}\, q^2 W(q);\; B(r)\simeq -
\frac{1}{3} \int d{\bf q} \, q^2 \tilde W'(q);\;C(r )\simeq 0.
\label{DC23b}
\end{eqnarray}

Practically the approximation $C(r)\simeq 0$ is always applicable
(see below Sec.III) and the general kinetic equation (\ref{DC16b})
for the Fourier-transform of the velocity distribution function
takes the form
\begin{eqnarray}
\frac{df_g({\bf r},t)}{dt} = A(r)f_g ({\bf r})+ B_\alpha
(r)\frac{\partial f_g ({\bf
r},t)}{\partial {\bf r}_\alpha} . \label{DC23c}
\end{eqnarray}

Then the simplified kinetic equation for the case of short-range
on $q$-variable PTF (non-equilibrium, in general case) reads
\begin{eqnarray}
\frac{df_g(r,t)}{dt} = A_0 r^2 f_g(r)+ B_0 r
\frac{\partial f_g (r)}{\partial r} ,\label{DC25b}
\end{eqnarray}
where $A_0\equiv -1/6 \int d{\bf q}\, q^2 W(q)$ and $B_0\equiv
-1/3 \int d{\bf q} \, q^2 \tilde W'(q)$.

The stationary solution of Eq.~(\ref{DC22b}) for $C(r)=0$ reads
\begin{eqnarray}
f_g(r,t)=C exp\,\left[-\int_0^r dr'\frac{A(r')}{r' B(r')}\right]
=C exp\,\left[-\frac{A_0 r^2}{2 B_0}\right] \label{DC26b}
\end{eqnarray}
The respective normalized stationary momentum distribution equals
\begin{eqnarray}
f_g(p)=\frac{N_g B_0^{3/2}}{(2 \pi A_0)^{3/2}}exp \,[-\frac{B_0
p^2}{2 A_0}]\label{DC28b}
\end{eqnarray}
Therefore in  Eq.~(\ref{DC26b}) the constant $C=N_g$. Equation
(\ref{DC25b}) and this distribution are the generalization of the
Fokker-Planck case for normal diffusion on non-equilibrium
situation, when the prescribed $ W({\bf q,p})$ is determined,
e.g., by some non-Maxwellian distribution of the small particles
$f_b$. To show this by other way let us make the Fourier
transformation of (\ref{DC16b}) with $C=0$ and the respective $A$
and $B_\alpha$:
\begin{eqnarray}
\frac{df_g({\bf p},t)}{dt} = - A_0 \frac{\partial^2 f_g({\bf p},t)}{\partial p^2}
- B_0 \frac{\partial (p_\alpha f_g
({\bf p},t))}{\partial p_\alpha} ,\label{DC29b}
\end{eqnarray}
Therefore we arrive to the Fokker-Planck type equation with the
friction coefficient $\beta\equiv-B_0$ and diffusion coefficient
$D=-A_0/M^2$. In general these coefficients (Eq.~(\ref{DC23b})) do
not satisfy to the Einstein relation.

In the case of equilibrium $W$-function (e.g., $f_b=f_b^0$, see
above) the equality $\tilde W'(q)=W(q)/2M T_b$ is fulfilled. Then with necessary accuracy (the second term in
Eq.~(\ref{DC23b}) with $W'$ is of order $\mu/M$ and negligible in comparison with the first one) we find
$A(r)/r B(r)\equiv A_0/B_0 =M T_b$. In this case the
Einstein relation between the diffusion and friction coefficients $D=\beta T/M$
exists and the standard Fokker-Planck equation is valid.

\section{The models of anomalous diffusion in $V$ - space}

Now we can calculate the coefficients for the models of anomalous
diffusion.

At first we calculate the simple model system of the hard spheres
with the different masses $m$ and $M\gg m$, $d\sigma/do=a^2/4$.
Let us suppose that in the model under consideration the small
particles are described by the prescribed stationary distribution
$f_b=n_b \phi_b /u_0^3$ (where $\phi_b$ is non-dimensional
distribution, $u_0$ is the characteristic velocity for the
distribution of the small particles) and $\xi \equiv (u^2+
v^2-{\bf q\cdot v} /\mu)/u_0^2$.
\begin{eqnarray}
W_a ({\bf q, p})= \frac{n_b a^2 \pi}{2\mu^2 u_0 q}
\int^\infty_{(q^2/4\mu^2+v^2-{\bf{q\cdot v}}/\mu)/u_0^2} d\xi
\,\cdot \phi _b (\xi).\label{DC30b}
\end{eqnarray}

If the distribution $\phi_b(\xi)=1/\xi^\gamma$ ($\gamma>1$)
possess a long-tail we get
\begin{eqnarray}
W_a({\bf q, p})= \frac{n_b a^2 \pi}{2\mu^2 u_0
q}\frac{\xi^{1-\gamma}}{(1-\gamma)}|_{\xi_0}^\infty= \frac{n_b a^2
\pi}{2\mu^2 u_0 q}\frac{\xi_0^{1-\gamma}}{(\gamma-1)}
,\label{DC31b}
\end{eqnarray}
where $\xi_0\equiv (q^2/4\mu^2+v^2-{\bf{q\cdot v}}/\mu)/u_0^2$.

For the case $p=0$ the value $\xi_0 \rightarrow \tilde \xi_0
\equiv q^2/4\mu^2 u_0^2$ and we arrive to the expression for
anomalous $W \equiv W_a$
\begin{eqnarray}
W_a ({\bf q, p=0})=\frac{n_b a^2
\pi}{2^{3-2\gamma}(\gamma-1)\mu^{4-2\gamma}
u_0^{3-2\gamma}q^{2\gamma-1}}\equiv \frac{C_a}{q^{2\gamma-1}}
.\label{DC32b}
\end{eqnarray}

To determine the structure of the transport process and the
kinetic equation in the velocity space we have find also the functions
$\tilde W'(q)$ and $\tilde W''(q)$.

If $p\neq 0$ to find $\tilde W'(q)$ and $\tilde W''(q)$ we have
use the full value $\xi_0\equiv (q^2/4\mu^2+p^2/M^2-{\bf{q\cdot
p}}/M \mu)/u_0^2$ and it derivatives on ${\bf q \cdot p}$ at
$p=0$, $\xi'_0=-1/M \mu u_0^2$ and $\xi''_0=0$. Then
\begin{eqnarray}
\tilde W'({\bf q, p})\equiv \frac{n_b a^2 \pi}{2M \mu^3 u^3_0
q}\xi_0^{-\gamma};
 \;\; \; \tilde W''({\bf q, p})\equiv \frac{n_b
a^2 \pi\gamma}{2M^2 \mu^4 u^5_0 q}\xi_0^{-\gamma-1} \label{DC34b}
\end{eqnarray}
Therefore for $p=0$ ($\xi_0 \rightarrow \tilde \xi_0$) we obtain
the functions
\begin{eqnarray}
\tilde W'(q)\equiv \frac{(4 \mu^2u_0^2)^\gamma n_b a^2 \pi}{2M
\mu^3 u^3_0 q^{2\gamma+1}};
 \;\; \; \tilde W''(q)\equiv \frac{(4 \mu^2u_0^2)^{\gamma+1} n_b
a^2 \pi\gamma}{2M^2 \mu^4 u^5_0 q^{2\gamma+3}} \label{DC35b}
\end{eqnarray}

The function $A(r)$ according to Eq.~(\ref{DC17b})
\begin{eqnarray}
A(r)\equiv 4\pi \int_0^\infty dq q^2 \left[\frac{sin\, (q
r)}{qr}-1\right]W(q)= 4\pi C_a
 \int_0^\infty d q
\frac{1}{q^{2\gamma-3}} \left[\frac {sin(qr)}{qr}-1\right]
\label{DC33b}
\end{eqnarray}

Comparing the reduced equation (see below) in the velocity space
with the diffusion in coordinate space
($2\gamma-1\leftrightarrow\alpha$ and $W(q)=C/q^{2\gamma-1}$) we
can establish that the convergence of the integral in the right
side of Eq.~(\ref{DC33b}) (3d case) is provided if $3<2\gamma-1<5$
or $2<\gamma<3$. The inequality $\gamma<3$ provides the
convergence for small $q$ ($q\rightarrow0$) and the inequality
$\gamma>2$ provides the convergence for $q\rightarrow\infty$.

We have establish now the conditions of convergence the integrals
for $B(r)$ and $C(r)$
\begin{eqnarray}
B(r)= \frac{4\pi}{r^2} \int_0^\infty dq q^2 \left[cos\, (q
r)-\frac{sin (q r)}{q r}\right]\tilde W'(q) \label{DC36b}
\end{eqnarray}

Convergence $B(r)$ exists for small $q$ if $\gamma<2$ and for
large $q\rightarrow\infty$ for $\gamma>1/2$.

Finally for $C(r)$ convergence is determined by the equalities
$\gamma<2$ for small $q$ and $\gamma>1$ for large $q$
\begin{eqnarray}
C(r )=\frac{2\pi}{r^2} \int_0^\infty dq q^4 \left[\frac{2 sin (q
r)}{q^3 r^3}-\frac{2 cos\, (q r)}{q^2 r^2}-\frac{sin
(qr)}{qr}+\frac{1}{3}\right]\tilde W''(q) \label{DC37b}
\end{eqnarray}

Therefore to provide convergence for $A$, $B$, $C$ for large $q$
we have provide convergence for $A$, that means $\gamma>2$. To
provide convergence for small $q$ it is enough to provide
convergence for $B$ and $C$, that means $\gamma<2$. Therefore for
the purely power behavior of the function $f_b(\xi)$ convergence
is absent. However, for existence of the anomalous diffusion in
the momentum space in reality the convergence for small $q$ is
always provided, e.g., by finite value of $v$ or by change of the
small $q$-behavior of $W(q)$ (compare with the examples of
anomalous diffusion in coordinate space [2]). Therefore, in the
model under consideration, the "anomalous diffusion in velocity
space"  for the power behavior of $W(q)$, $\tilde W'(q)$ and
$\tilde W''(q)$ on large $q$ exists if for large $q$ the
asymptotic behavior of $W(q\rightarrow\infty)\sim 1/q^{2\gamma-1}$
with $\gamma>2$. At the same time the expansion of the exponential
function in Eqs. (\ref{DC17b})-(\ref{DC21b}) under the integrals,
which leads to the Fokker-Planck type kinetic equation is invalid
for the power-type kernels $W(\bf {q, p)}$.

Let us consider now the formal general model for which we will not
connect the functions $W(q)$, $\tilde W'(q)$ and $\tilde W''(q)$
with the concrete form of $W({\bf q, p})$. In this case one can
suggest that the functions possess the independent one from
another power-type $q$-dependence.

As an example, this dependence can be taken as the power type for
three functions $W(q)\equiv a / q^\alpha$, $\tilde W'(q)\equiv b
/q^\beta$ and $\tilde W''(q)\equiv c /q^\eta$ , where $\alpha$,
$\beta$ and $\eta$ are independent and positive. Then as follows
from the consideration above the convergence of the function $W$
exists if \,$5>\alpha>3$ (for asymptotically small and large $q$
respectively). For the function $\tilde W'(q)$ the convergence
condition is $5>\beta>2$ for asymptotically small and large $q$
respectively.
Finally for the function $\tilde W''(q)$ the
convergence condition is $7>\eta>5$ (for asymptotically small and
large $q$ respectively).

For this example the kinetic equation Eq.~(\ref{DC16b}) reads
\begin{eqnarray}
\frac{df_g({\bf r},t)}{dt} = P_0 r^{\alpha-3} f({\bf r},t)+
r^{\beta-5} P_1 r_i\frac{\partial}{\partial r_i} f({\bf r},t)+
r^{\eta-7}P_2 r_i r_j \frac{\partial^2}{\partial r_i
\partial r_j}f({\bf r},t), \label{DC38b}
\end{eqnarray}
where
\begin{eqnarray}
P_0= 4\pi a \int_0^\infty d\zeta \zeta^{2-\alpha}
\left[\frac{sin\, \zeta}{\zeta}-1\right] \label{DC39b}
\end{eqnarray}
\begin{eqnarray}
P_1= 4\pi b \int_0^\infty d\zeta q\zeta^{2-\beta} \left[cos\,
\zeta-\frac{sin \zeta}{\zeta}\right] \label{DC40b}
\end{eqnarray}
\begin{eqnarray}
P_2=4\pi c \int_0^\infty d\zeta \zeta^{4-\eta} \left[\frac{ sin
\zeta}{\zeta^3}-\frac{ cos\,\zeta}{\zeta^2}-\frac{sin
\zeta}{2\zeta}+\frac{1}{6}\right] \label{DC41b}
\end{eqnarray}

Taking into account the isotropy in $r$-space we can
rewrite Eq.~(\ref{DC38b}) in the form
\begin{eqnarray}
\frac{df_g(r,t)}{dt} = P_0 r^{\alpha-3} f(r)+ r^{\beta-4} P_1
\frac{\partial}{\partial r} f(r,t)+ r^{\eta-5}P_2
\frac{\partial^2}{\partial r^2}f(r,t), \label{DC42b}
\end{eqnarray}

Naturally, Eqs.~(\ref{DC38b}),(\ref{DC42b}) can be formally
rewritten in momentum (or in velocity) space via the fractional
derivatives of various orders. Therefore, as is easy to see, for the
purely power behavior of the functions  $W(q)$, $\tilde W'(q)$ and
$\tilde W''(q)$ the solution with the convergent coefficients exists for
the powers in the intervals mentioned above.
The universal type of anomalous diffusion in velocity space in the case under consideration,
therefore, exists if $5>\alpha>3$, $5>\beta>2$ and $7>\eta>5$.
It appears even for the cases, when the functions $W(q)$, $\tilde
W'(q)$ and $\tilde W''(q)$ have not the short-range cutting. Of cause the general description is
also valid for the more complicated functions $W$, $W'$ and $W''$ possessing the non-power short
range parts.

Now let us take into account the important circumstance: usually in the problem under consideration there is
a small parameter $~\mu/M$, which can simplify description of the velocity diffusion.
As is easy to see, e.g., on the basis
of the particular cases (e.g., Eq.~(\ref{DC35b})) for the convergent kernels of anomalous transport the term
with second space derivative in general equations for distributions  Eqs.~(\ref{DC16b}),(\ref{DC22b}) is small
in comparison with the term with the second derivative (as well as for the case of normal diffusion in velocity space).
This smallness is of the order of the small ratio $\mu/M$ of the mass of the particles. Therefore for the most physically
important kernels, describing the anomalous velocity diffusion the term with the second space derivative can be omitted
and for non-stationary anisotropic and isotropic cases the diffusion equation respectively reads
\begin{eqnarray}
\frac{df_g({\bf r},t)}{dt} = A(r)f_g ({\bf r})+ B_\alpha
(r)\frac{\partial f_g ({\bf
r},t)}{\partial{\bf r}_\alpha} \label{DC47a}
\end{eqnarray}
and
\begin{eqnarray}
\frac{df_g(r,t)}{dt} = A (r) f_g(r)+ B(r)r \frac{\partial f_g(r,t)}{\partial
r}  \label{DC47b}
\end{eqnarray}
For the case of purely power behavior $W(q)=1/q^\alpha$ and $W'(q)=1/q^\beta$ we have, as above, $A(r)=P_0 r^{\alpha-3}$ and $rB(r)=P_1 r^{\beta-4}$ (with inequalities $5>\alpha>3$ and $5>\beta>2$). The stationary solution of Eq. (\ref{DC47b})) (see, also (\ref{DC26b})) for the case under consideration reads
\begin{eqnarray}
f^{St}_g (r)= C exp\,[-\int^r dr'\frac{A(r')}{r' B(r')}]=C exp\,[-\frac{P_0 r^{\alpha-\beta+2}}{P_1 (\alpha-\beta+2)}]
\label{DC47b1}
\end{eqnarray}
where $5>\alpha-\beta+2>0$.

To find the solution in isotropic non-stationary case Eq.~(\ref{DC47b}) has to be written in the form
\begin{eqnarray}
\frac{dX (r,t)}{dt}-B(r)r \frac{\partial}{\partial
r} X (r,t) = A (r), \label{DC48b}
\end{eqnarray}
where $X(r,t)\equiv ln f_g(r,t)$. The general non-stationary
solution of this equation can be written as the sum of general
solution of the homogeneous equation $Y(r,t)$ (Eq.~(\ref{DC48b})
in which $A(r)$ is taken equals zero)
\begin{eqnarray}
Y (r,t) = \Phi (\xi), \,\, \xi \equiv t+\int^r_{r_0} dr'\frac{1}{r' B(r')},  \label{DC49b}
\end{eqnarray}
where $\Phi$ is the arbitrary function and the particular solution $Z(r,t)$ of the non-homogeneous equation Eq.~(\ref{DC48b}):
\begin{eqnarray}
Z (r,t)\equiv f^{St}_g (r)= -\int^r_{r_0} dr'\frac{A(r')}{r' B(r')}
\label{DC50b}
\end{eqnarray}
Therefore
\begin{eqnarray}
f_g(r,t) = exp\, [X(r,t)] \equiv exp \left[ Y+Z \right]=L(\xi)f^{St}_g (r), \label{DC51b}
\end{eqnarray}
where $L(\xi)$ is the arbitrary function of $\xi$, which has to be found from the initial condition $f_g(r,t=0)\equiv \phi_0(r)$.

The variable $\xi (r,t)$ equals
\begin{eqnarray}
\xi (r,t) =t+ \int^r_{r_0} dr'\frac{1}{r' B(r')}=t+\frac{r^{5-\beta}}{P_1 (5-\beta)}+c,  \label{DC53b}
\end{eqnarray}
where $c$ is the arbitrary constant, which can be omitted due to presence of the arbitrary
function $L$ and the values $5>\alpha-\beta+2>0$, $3>5-\beta>0$.
The general non-stationary solution for the case under consideration reads
\begin{eqnarray}
f_g(r,t)=L \left(t+\frac{r^{5-\beta}}{P_1 (5-\beta)}\right) \, exp \left[-\frac{P_0 r^{\alpha-\beta+2}}{P_1 (\alpha-\beta+2)}\right], \label{DC54b}
\end{eqnarray}
The unknown function $L$ can be found from Eq.~(\ref{DC54b}) and the initial condition $f_g(r,0)\equiv \phi_g(r)$:
\begin{eqnarray}
L \left(\frac{r^{5-\beta}}{P_1 (5-\beta)}\right) \, exp \left[-\frac{P_0 r^{\alpha-\beta+2}}{P_1 (\alpha-\beta+2)}\right]=\phi_g(r), \label{DC55b}
\end{eqnarray}
The function $\phi_g(r)=\int d^3 p exp (i{\bf p r}) f_g(p, t=0)$ is the Fourier-component of the initial distribution in momentum space.
By use the notation $\zeta\equiv r^{5-\beta}/[P_1 (5-\beta)]$ (what means $r(\zeta)\equiv[P_1(5-\beta)\zeta]^{1/(5-\beta)}$) we find
\begin{eqnarray}
L(\zeta) =\phi_g[r(\zeta)]\, exp \left\{\frac{P_0 [r(\zeta)]^{\alpha-\beta+2}}{P_1 (\alpha-\beta+2)}\right\}, \label{DC55b}
\end{eqnarray}
Therefore the time-dependent solution is equal
\begin{eqnarray}
f_g(r,t)=\phi_g[r(\zeta+t)]\, exp \left\{\frac{P_0 [r(\zeta+t)]^{\alpha-\beta+2}}{P_1 (\alpha-\beta+2)}\right\}\,
exp \left[-\frac{P_0 r^{\alpha-\beta+2}}{P_1 (\alpha-\beta+2)}\right], \label{DC56b}
\end{eqnarray}
where we have express $r(\zeta+t)\equiv[P_1(5-\beta)(\zeta+t)]^{1/(5-\beta)}$ as a function of $r,t$:
\begin{eqnarray}
r(\zeta+t)\equiv[P_1(5-\beta)(\zeta+t)]^{1/(5-\beta)}\equiv[r^{5-\beta}+P_1(5-\beta)t]^{1/(5-\beta)}, \label{DC57b}
\end{eqnarray}
or finally

\begin{eqnarray}
f_g(r,t)=\phi_g\left([r^{5-\beta}+P_1(5-\beta)\,t]^{1/(5-\beta)}\right)\, exp \left\{\frac{P_0\,
[r^{5-\beta}+P_1\,(5-\beta)\,t]^{(\alpha-\beta+2)/(5-\beta)}-P_0 \,r^{\alpha-\beta+2}}{P_1\, (\alpha-\beta+2)}\right\}. \label{DC56b}
\end{eqnarray}
It is necessary to stress that for the fractional powers $1/(5-\beta)$ and (or) $(\alpha-\beta+2)/(5-\beta)$ the real
solution exists only if $P_1>0$. The limit $t\rightarrow \infty$ for the solution only for the specific initial conditions
can coincide with the stationary solution.

For the power dependence of the functions $W(q)$ and $W'(q)$ the equation (\ref{DC47a}) can be formally written in fractional derivatives:
\begin{eqnarray}
\frac{df_g({\bf p},t)}{dt} = P_0 D^\nu f_g({\bf p},t)-P_1 (3+\gamma)D^\gamma f_g({\bf p},t)+P_1 p_\alpha D_\alpha^{\gamma+1}
f_g ({\bf p},t), \label{DC56b}
\end{eqnarray}
where $\nu\equiv \alpha-3$, $\gamma\equiv \beta-5$ and $D_\alpha^{\gamma+1}f_g ({\bf p},t)
\equiv i \int d^3r exp (-i{\bf pr}) )r_\alpha r^\gamma f_g ({\bf r},t)$.

The specific case of anomalous diffusion in velocity space, which
leads to the equation similar to one in [14,15] is derived in the
Appendix on the basis of general equation (\ref{DC16b}).

\section{The model of diffusion on basis of the Boltzmann collisions with drift distribution}

Let us consider the simplest case of non-equilibrium, but
stationary distribution $f_b$, namely the shifted velocity
distribution.

The evident generalization of the PT-function for the case of the
shifted velocity distribution of the light particles (with the
drift velocity ${\bf u_d}$) the PT-function $w_B^d({\bf q, p})$
(${\bf p}=M {\bf v}$) reads
\begin{eqnarray}
w_B^d({\bf q, p, u_d})= \frac{2\pi}{\mu^2 q} \int_{q /2\mu}^\infty d u
u\,\cdot \frac{d \sigma}{do} \left[\arccos \, (1-\frac{q^2}{2
\mu^2 u^2}), u \right] \times \nonumber\\
f_b (u^2+ ({\bf v-u_d})^2-{\bf q} \cdot {\bf (v-u_d)} /\mu)
.\label{DB1}
\end{eqnarray}
Again as in the Section II to find the coefficients in the
kinetic equation let us use the way, based on the difference
between the velocities of the light and heavy particles. At the same time the driven velocity $u_d$
is not, generally speaking, small in comparison with the current characteristic velocities
$u$ and $q/\mu$ of the small particles.

For calculation of the function $A_\alpha$ we have take into
account that in general the scalar function $w_B^d({\bf q, p,
u_d})$ has to be taken in the form $W({\bf q,p,u_d})=W(q,p,u_d,
l,\xi,\eta)$ (here $l \equiv ({\bf q \cdot p_d}$)) and expanded on
$\xi \equiv ({\bf q\cdot p})$ and $\eta\equiv (M {\bf u_d \cdot
p})\equiv ({\bf p_d \cdot p})$. In fact it is the expansion on
velocity ${\bf v}$, which is small in comparison with other
characteristic velocities $q/\mu$, $u$ and $u_d$. As we showed
above (for the case $u_d=0$) to arrive to the simple and solvable
equation for the distribution $f_g$, taking into account smallness
of $v$ in comparison with the characteristic velocities and $v^2$
in comparison with $({\bf q \cdot v}/\mu$)) we have approximate
the function $W(q,p,u_d, l,\xi,\eta) \simeq W(q,u_d, l,\xi,\eta)$,
because we are interested mainly the high value of $q$ for
anomalous transport. At the same time after this type of
neglecting and expansion on $\xi$ and (for $u_d\neq 0$) $\eta$ we
can arrive for the special case of the kernels (e.g., purely
power-type kernels on $q$, which are often considering for
diffusion in coordinate space) to the divergence in some
coefficients of the diffusion equation created by the region of a
small $q$. This divergence is really absent for the realistic PT
functions, which have cutting at small $q$. This cutting for small
$q$ has the physical reasons and is not related with the
neglecting $v=p/M$ in the approximation $W({\bf q,p,u_d})\equiv
W(q,p,u_d, l,\xi,\eta) \simeq W(q,u_d, l,\xi,\eta)$.

Let us expand $W(q,p, u_d, l,\xi,\eta)$:
\begin{eqnarray}
W({\bf q,p,u_d})= W(q,p, u_d, l,\xi,\eta)\simeq
W_0(q,p, u_d,l)+\partial W/\partial \xi\mid_{\xi, \eta=0} \xi +
\partial W/\partial \eta\mid_{\xi, \eta=0}  \eta
+\nonumber\\
\frac {1}{2}\partial^2 W/\partial \xi^2 \mid_{\xi, \eta=0}
\xi^2+\frac {1}{2}\partial^2 W/\partial \eta^2 \mid_{\xi, \eta=0}
\eta^2+
\partial^2 W/\partial \xi \partial \eta \mid_{\xi, \eta=0}\xi \eta
,\label{DB2}
\end{eqnarray}
where $W_0(q,p,u_d,l)\equiv W(q,p, u_d, l,\xi,\eta)\mid_{\xi, \eta=0}$.
Then introducing $V_1(q,p, u_d,l)=\partial W/\partial \xi\mid_{\xi,
\eta=0}, U_1(q,p,u_d,l)=\partial W/\partial \eta\mid_{\xi, \eta=0},
V_2(q,p,u_d,l)=\frac {1}{2}\partial^2 W/\partial \xi^2 \mid_{\xi,
\eta=0}, U_2(q,p,u_d,l)=\frac {1}{2}\partial^2 W/\partial \eta^2
\mid_{\xi, \eta=0}, W_2(q,p,u_d,l)=\partial^2 W/\partial \xi
\partial \eta \mid_{\xi, \eta=0}$ we can rewrite Eq.~(\ref{DB2})
in the form
\begin{eqnarray}
W(q,p, u_d, l,\xi,\eta) \simeq W_0(q,p,u_d,l)+V_1 ({\bf q\cdot p}) +\nonumber\\
U_1 ({\bf p_d \cdot p}) + V_2 ({\bf q\cdot p})^2+U_2 ({\bf p_d
\cdot p})^2+ W_2 ({\bf q\cdot p}) ({\bf p_d \cdot p})\simeq \nonumber\\ W_0(q,p=0,u_d,l)+V_1(q,p=0,u_d,l) ({\bf q\cdot p})+
U_1(q,p=0,u_d,l) ({\bf p_d \cdot p}) , \label{DB3}
\end{eqnarray}
Finally we put $p=0$ in the coefficients of Eq.~(\ref{DB3}) and omit the terms with second order derivatives due to existence of the small parameter $~\mu/M$. It means only the terms of order $\sqrt{\mu/M}$ are essential in the expansion of $W({\bf q,p,u_d})$.
Let us calculate $W ({\bf q, p+q,u_d})$ taking into account the difference
between the values of the characteristic momenta ${\bf q, p+q,u_d}$ or
velocities.

Then the expansion for the function $W ({\bf q, p+q,u_d})$ with the necessary accuracy reads
\begin{eqnarray}
W ({\bf q, p+q, u_d})\simeq W ({\bf q, p, u_d})+\{q_{\alpha}
\partial /\partial p_{\alpha}+\frac{1}{2}q_\alpha q_{\beta}
\frac{\partial^2 }{\partial p_ {\alpha} \partial p_{\beta}}
\}W({\bf q, p,u_d})\simeq
\nonumber\\\\W_0(q,p=0,u_d,l)+V_1(q,p=0,u_d,l) ({\bf q\cdot p})+
U_1(q,p=0,u_d,l) ({\bf p_d \cdot p})+\nonumber\\
V_1(q,p=0,u_d,l) {\bf q}^2+
U_1(q,p=0,u_d,l) ({\bf q \cdot  p_d})
\label{DB3a}
\end{eqnarray}
or
\begin{eqnarray}
W ({\bf q, p+q, u_d})\simeq W_0(q,p=0,u_d,l)+\nonumber\\ V_1(q,p=0,u_d,l) [{\bf q\cdot (p+q)}]+
U_1(q,p=0,u_d,l) [{\bf p_d \cdot (p+q)}].
\label{DB4}
\end{eqnarray}
Then the kinetic equation reads
\begin{eqnarray}
\frac{df_g({\bf p},t)}{dt} = \int d{\bf q} \{[W_0(q,p=0,u_d,l)+V_1(q,p=0,u_d,l) ({\bf q\cdot p})+
U_1(q,p=0,u_d,l) ({\bf p_d \cdot p})\nonumber\\+
V_1(q,p=0,u_d,l) {\bf q}^2+U_1(q,p=0,u_d,l)({\bf q \cdot  p_d})]\nonumber\\ f_g({\bf p+q},t)- [W_0(q,p=0,u_d,l)+V_1(q,p=0,u_d,l) ({\bf q\cdot p})+\nonumber\\
U_1(q,p=0,u_d,l) ({\bf p_d \cdot p})] f_g({\bf p}, t)\}.
\label{DB5}
\end{eqnarray}
After the Fourier-transformation $f_g ({\bf r})=\int \frac{d{\bf
p}}{(2\pi)^3} exp(i{\bf p r})f_g ({\bf p},t)$ Eq.~(\ref{DC13b})
reads:
\begin{eqnarray}
\frac{df_g({\bf r},t)}{dt}= \int d{\bf q}\{exp(-i{\bf q
r)}[W_0(q,p=0,u_d,l)-i V_1(q,p=0,u_d,l) ({\bf
q\cdot\frac{\partial}{\partial{\bf r}}})\nonumber\\- i
U_1(q,p=0,u_d,l) ({\bf p_d \cdot \frac{\partial}{\partial{\bf r}}
})]f_g({\bf r},t))\nonumber\\ - [W_0(q,p=0,u_d,l)-i
V_1(q,p=0,u_d,l) ({\bf q\cdot \frac{\partial}{\partial{\bf r}}})-
i U_1(q,p=0,u_d,l) ({\bf p_d \cdot \frac{\partial}{\partial{\bf r}}})]f_g({\bf r},t))\}\nonumber\\
 \label{DB6}
\end{eqnarray}
Therefore
\begin{eqnarray}
\frac{df_g({\bf r},t)}{dt}=A_d({\bf r})f_g({\bf r},t)+{\bf B}_d({\bf r})\frac{\partial}{\partial{\bf r}}f_g({\bf r},t),
\label{DB7}
\end{eqnarray}
where
\begin{eqnarray}
A_d({\bf r, p_d})=\int d{\bf q}[exp(-i{\bf q
r)}-1]W_0(q,p=0,u_d,l),
\label{DB7}
\end{eqnarray}
\begin{eqnarray}
{\bf B}_d({\bf r,p_d})=-i \int d{\bf q}[exp(-i{\bf q
r)}-1]\{V_1(q,p=0,u_d,l) {\bf q}+U_1(q,p=0,u_d,l){\bf p}_d\}\equiv\nonumber\\{\bf r}B'_d({\bf r,p_d})+{\bf p_d} B''_d({\bf r,p_d}),
\label{DB8}
\end{eqnarray}

In more detail we consider this equation in the separate paper.

\section{Appendix. Anomalous velocity diffusion for the specific case
B(r)=const, C(r)=0}

Let us consider now formally the specific particular case of anomalous diffusion,
when the specific structure of the PTF $W({\bf q, p})$ provides a
rapid (let say, exponential) decrease of the functions $\tilde
W'(q)$ and $\tilde W''(q)$. Therefore the exponential function
under the integrals in the coefficients $B(r)$ and $C(r)$ can be
expanded, that means $B(r)=B_0$ and $C(r)\simeq 0$ respectively.
At the same time the function $W(q)\equiv a/q^\alpha$ has a purely
power dependence on $q$.

Then the kinetic equation Eq.~(\ref{DC16b}) reads
\begin{eqnarray}
\frac{df_g({\bf r},t)}{dt} = P_0 r^{\alpha-3} f_g({\bf r},t)+ B_0
r_i\frac{\partial}{\partial r_i} f_g({\bf r},t), \label{DC43b}
\end{eqnarray}
or formally in the momentum space
\begin{eqnarray}
\frac{df_g({\bf p},t)}{dt} = P_0 D^\nu f_g({\bf p},t)- B_0
\frac{\partial}{\partial p_i} [p_i f_g({\bf p},t)], \label{DC44b}
\end{eqnarray}
where $\nu \equiv(\alpha-3)$ ($2>\nu>0$) and we introduced the
fractional differentiation operator $D^\nu f({\bf p},t)\equiv \int
d{\bf r} r^\nu exp(-i{\bf pr}) f({\bf r},t) $ in the momentum
space to compare this equation with the similar one in [14]. The
stationary solution of Eq.~(\ref{DC43b}) is equal
\begin{eqnarray}
f_g(r) = C exp\;[-\frac {P_0 r^{\nu-1}}{B_0}] \label{DC45b}
\end{eqnarray}
\begin{eqnarray}
f_g(p) = C \int d^3 r exp(-i{\bf pr})exp\;[-\frac {P_0 r^{\nu-1}}{B_0}]\equiv
\frac{4\pi C}{p} \int^\infty_0  dr r sin (pr) exp\;[-\frac {P_0 r^{\nu-1}}{B_0}]\label{DC46b}
\end{eqnarray}
For the case $\nu=1$ we find $f({\bf p})=n_g \delta({\bf p})$,
therefore $C=n_g exp\;[P_0/B_0]/(2\pi)^3$. The similar
consideration has to be used for other types of anomalous
diffusion in velocity space. The physically important applications
based on the physical models for $W({\bf q, p})$ function will be
considered separately.

\section{Conclusions}
In this paper the problem of anomalous diffusion in momentum
(velocity) space is consequently considered. The new kinetic
equation for anomalous diffusion in velocity space is derived
without suggestion about existence of the equilibrium stationary
distribution function. Namely for the strongly non-equilibrium
situations the long tails in PT-functions can manifest themselves.
The model of anomalous diffusion in velocity space is described on
the basis of the respective expansion of the kernel in master
equation. The conditions of the convergence for the coefficients
of the kinetic equation are found for the particular cases. The
wide variety of the anomalous processes in velocity space exists,
because even in isotropic case the three different coefficients in
the general diffusion equation are present (one of them is usually
negligible). The example of the Boltzmann kernel with the
prescribed distribution function for the light particles is
studied, in particular for the hard spheres interaction. In
general the Einstein relation for such situation is not
applicable, because the stationary state can be far from
equilibrium. For the normal diffusion the friction and diffusion
coefficient are explicitly found for the non-equilibrium case. For
equilibrium case the usual Fokker-Plank equation is reproduced as
the particular case.

The non-stationary and in general non-equilibrium for
$t\rightarrow 0$ solution is found for the definite initial
conditions.

The kinetic equation for the heavy particles distribution in the
case of the prescribed distribution function for the light
particles, which possess a drift velocity is derived in the
suggested general approach.

\section*{Acknowledgment}
The author is thankful to E. Allahyarov, W. Ebeling, M.Yu.
Romanovsky and I.M. Sokolov for valuable discussions of some
problems, reflected in this work and E. Barkai for the useful
remark. I also express my gratitude to the Netherlands
Organization for Scientific Research (grant NWO 047.017.2006.007)
and the Russian Foundation for Basic Research for support of my
investigations on the problems of stochastic transport.

\end{document}